\documentclass[preprint,aps,prd,showpacs,nofootinbib]{revtex4}
\usepackage{mathrsfs}
\usepackage{amsmath}
\usepackage{graphicx}
\usepackage{subfigure}

\newcommand{\bea}{\begin{eqnarray}}
\newcommand{\eea}{\end{eqnarray}}
\newcommand{\beq}{\begin{equation}}
\newcommand{\eeq}{\end{equation}}

\newcommand{\cev}[1]{\reflectbox{\ensuremath{\vec{\reflectbox{\ensuremath{#1}}}}}}

\def\/{\over}

\begin{document}
\title{\bf  Can spacetime curvature induced corrections to Lamb shift be observable?}
\author{  Wenting Zhou $^{1}$ and  Hongwei Yu $^{1,2,}$\footnote{Corresponding author}}
\affiliation{ $^1$ Institute of Physics and Key Laboratory of Low
Dimensional Quantum Structures and Quantum Control of Ministry of
Education, Hunan Normal University, Changsha, Hunan 410081, China\\
$^2$ Center for Nonlinear Science and Department of Physics, Ningbo
University, Ningbo, Zhejiang 315211, China}

\begin{abstract}
The Lamb shift results from the coupling of an atom to vacuum fluctuations of quantum fields, so corrections are expected to arise when the spacetime is curved since the vacuum fluctuations are modified by the presence of spacetime curvature. Here, we calculate the curvature-induced correction to the Lamb shift outside a spherically symmetric object and demonstrate that this correction can be remarkably significant outside a compact massive  astrophysical body. For instance, for a neutron star or a stellar mass black hole, the correction is $\sim$ 25\% at a
radial distance of $4GM/c^2$, $\sim$ 16\% at $10GM/c^2$ and as large as $\sim$ 1.6\% even at $100GM/c^2$, where $M$ is the mass of the object, $G$ the Newtonian constant, and $c$ the speed of light. In principle, we can look at the spectra from a distant compact super-massive body to find such corrections.  Therefore, our results suggest a possible way of detecting fundamental quantum effects in astronomical observations.
\end{abstract}
\maketitle

\baselineskip=16pt

The interplay of quantum theory and general relativity has brought new insights into our understanding of these two fundamental theories which are pillars of modern physics. The Hawking radiation of black holes is one of the most striking phenomena that are revealed by our attempt to establish quantum field theory in curved spacetime. On one hand, as one of the quantum effects that are unique to curved spacetime, the Hawking radiation, which has been extensively examined in many different contexts,  has been considered as a `` Rosetta stone" to relate quantum theory, general relativity and thermodynamics and is expected to be a indispensable part of a yet-to-be-found full theory of quantum gravity;  on the other hand, it still remains a tremendous task to experimentally verify the Hawking effect as well as other quantum effects unique to curved spacetime, although a lot of efforts have been made to observe it in analogue systems~\cite{UH}.  In consideration of the lack of direct experiment test of the quantum effects unique to curved spacetime, here we take a different approach, i.e., we ask,  what are the corrections, caused by the fact that the spacetime is curved,  to quantum effects already existing in flat spacetime? and can these corrections be observable?

In this regard, the quantum effect that first comes to our mind is the Lamb shift, a subtle difference in energy between two levels of an atom which arises as a result of the coupling of the atom to vacuum fluctuations of quantum fields. The Lamb shift is one of the most important
and inspiring discoveries of the last century that marked  the beginning of modern quantum electromagnetic field theory. Since its discovery in experiment in 1947 \cite{Lamb 47}, the Lamb shift has attracted a great deal of attention and been measured with great precision. It is worthwhile to note that the measurement of the Lamb shift also plays an important role in our understanding of nuclear structure and in the determination of fundamental physical constants~\cite{Nature,Fischer,Niering,DBeauvoir,Schwob}. So far, on the theoretical front, the Lamb shift has been examined in various circumstances, for example, in the presence of
cavities~\cite{Meschede90}, or in a thermal bath~\cite{Barton72,Knight72,Farley and Wing81,ZhuYu09}. It has recently been shown that the non-inertial motion of the atom also
induces corrections to the Lamb shift~\cite{Audretsch95,Passante98,L.Rizzuto07,ZhuYu10}. However,
all the aforementioned studies are concerned with flat spacetimes. Therefore, it remains interesting to see what happens if the atom is placed in a curved spacetime rather than a flat
one. Now, a correction to the Lamb shift as opposed to its original value in a
flat spacetime is generally expected, since vacuum fluctuations of
quantum fields are modified by the presence of spacetime curvature~\footnote{Let us also note that the energy-level structure of an atom will in general be different from that in a flat space since it is now determined by the wave equation written in a curved space.}.
Here, we are concerned with the correction to the Lamb shift which is caused by the modified vacuum fluctuations due to the spacetime curvature outside a spherically symmetric astrophysical
object, and, surprisingly, we find that this correction is potentially observable, thus providing a possible way of checking fundamental quantum effects in astronomical observations.

The Lamb shift  has been studied in different physical contexts, all yielding the same
result in agreement with experiment with remarkable precision\cite{Feynman,Welton,Layzer,Erickson,Riis,Audretsch95}. Our calculation here is based upon the so-called Dalibard, Dupont-Roc, and Cohen-Tannoudji(DDC) formalism~\cite{Dalibard82,Dalibard84} in which the contribution of vacuum fluctuations to the energy-level shift of an atom and that of
the radiation reaction are distinctively separated. Since we are interested in the correction induced by the curvature, we consider, for simplicity, a two-level atom interacting with quantized
real massless scalar fields in vacuum at a fixed radial distance outside a spherical massive body. When a curved spacetime is concerned as opposed to a flat one, a delicate issue then arises as to how the vacuum state of the quantum fields is specified. In this paper, we assume that the scalar field is in the  Boulware vacuum~\cite{Boulware75}, since it is  the vacuum state outside a massive spherical
body which has not collapsed through its event horizon.
Two stationary states of the atom are represented by $|+\rangle$ and $|-\rangle$ respectively. The Hamiltonian $H$ of the system (atom$+$ bath of fluctuating scalar fields) is composed of the following three parts
 \beq
H_A(\tau)=\hbar\omega_0S_z(\tau)\;,
 \eeq
 \beq
H_F(\tau)=\int d^3 k\;\hbar\,\omega_{\vec{k}} a_{\vec{k}}^\dag
a_{\vec{k }}{dt\/d \tau}\;,
 \eeq
 \beq
H_I(\tau)={\mu}S_2(\tau)\phi(x(\tau))\;.\label{HI}
 \eeq
 Here $H_A(\tau)$, $H_F(\tau)$ and $H_I(\tau)$ are respectively the Hamiltonian
of the atom, the field and the Hamiltonian  describing the interaction between them.
$S_2(0)=\frac{1}{2}i(|-\rangle\langle+|-|+\rangle\langle-|)$, $S_z(0)
=\frac{1}{2}(|+\rangle\langle+|-|-\rangle\langle-|)$, and $\phi$ is the field operator.
 $a_{\vec{k}}^\dag$ and $a_{\vec{k }}$ are the field creation and annihilation operators.
  $\mu$ is the coupling constant which is assumed to be small.

Assuming the initial state of the atom is  $|b\rangle$ and the field
is in the Boulware vacuum, one can show, in the framework of
Dalibard, Dupont-Roc, and Cohen-Tannoudji
(DDC)~\cite{Dalibard82,Dalibard84, Audretsch95}, that the
contributions of vacuum fluctuations and radiation reaction to the
energy shift of level $b$ are respectively given by,
 \bea
(\delta E_b)_{vf}&=&-\frac{{i\mu^2}}{\hbar}\int^\tau_{\tau_0}d\tau'
C^F(x(\tau),x(\tau'))(\chi^A)_b(\tau,\tau')\;,\label{Evf}\\
(\delta E_b)_{rr}&=&-\frac{{i\mu^2}}{\hbar}\int^\tau_{\tau_0}d\tau'
\chi^F(x(\tau),x(\tau'))(C^A)_b(\tau,\tau')\;,\label{Err}
 \eea
where $\tau$ is the proper time of the atom and $x(\tau)$ represents its stationary trajectory which we will specify later.
$C^F$ and $\chi^F$ are the symmetric correlation function and linear susceptibility of
the field defined as
 \bea
 C^{F}(x(\tau),x(\tau')) &=& {1\/2}{\langle} 0| \{ \phi^f
(x(\tau)), \phi^f(x(\tau')) \} | 0 \rangle\;, \label{general form of
Cf}\\
 \chi^F(x(\tau),x(\tau')) &=& {1\/2}{\langle} 0| [
\phi^f(x(\tau)),\phi^f (x(\tau'))] | 0 \rangle\;,\label{general
form of Xf}
 \eea
and $(C^A)_b$ and $(\chi^A)_b$ are those of the atom, which are
given by
 \bea
(C^A)_{b}(\tau,\tau')&=&\frac{1}{2}\sum_{d}|\langle
b|S_2(0)|d\rangle|^2
   (e^{i\omega_{bd}\Delta\tau}+e^{-i\omega_{bd}\Delta\tau})\;,\\
(\chi^A)_{b}(\tau,\tau')&=&\frac{1}{2}\sum_{d}|\langle
b|S_2(0)|d\rangle|^2
   (e^{i\omega_{bd}\Delta\tau}-e^{-i\omega_{bd}\Delta\tau})\;.
 \eea
Here $\omega_{bd}=\omega_b-\omega_d$ is the energy spacing between
the two levels, $|\omega_{bd}|=\omega_0$ for $b\neq d$, $\Delta
\tau=\tau-\tau'$ and the sum over $d$ extends over a complete set of
states of the atom.

Now let us calculate the energy-level shift of a two-level atom at a fixed radial distance, of which the trajectory is described by
 \beq\label{traj} t(\tau)={1\/\sqrt{g_{00}}}(\tau-\tau_0),\ \ \
r(\tau)=r,\ \ \ \theta(\tau)=\theta,\ \ \ \phi(\tau)=\phi\;.
 \eeq
 First, we need
to solve the  Klein-Gordon equation in the given curved background, define the
Boulware vacuum and calculate the statistical functions of the field.
For this purpose, let us note that the metric of the spacetime outside
a spherical object  can be written in the Schwarzschild coordinates as
 \beq
ds^2= c^2\bigg(1-{r_s\over {r}}\bigg)\;dt^2-\bigg(1-{r_s\over
{r}}\bigg)^{-1}\;dr^2-r^2\,(d\theta^2+\sin^2\theta\,d\varphi^2)\;.
 \eeq
Here $r_s$ is the Schwarzschild radius, $r_s=2GM/c^2$ in which $M$
is the mass of the object, $G$ is  the gravitational constant and $c$
the speed of light.
Solving the Klein-Gordon equation satisfied by a massless scalar
field \cite{Dewitt75}, one finds, in the exterior region of the massive body, a complete set of normalized basis
functions
 \bea
 \vec{u}_{\omega
lm}=\sqrt{\frac{\hbar}{4\pi\omega c}}e^{-i\omega
t}\vec{R}_l(\omega|r)\;Y_{lm}(\theta,\varphi)\;,\label{outgoing modes}\\
\cev{u}_{\omega lm}=\sqrt{\frac{\hbar}{4\pi\omega c}}e^{-i\omega
t}\cev{R_l}(\omega|r)\;Y_{lm}(\theta,\varphi)\label{ingoing modes}\;,
 \eea
where $Y_{lm}(\theta,\varphi)$ are the spherical harmonics and the
radial functions satisfy the following differential equation
 \beq
\frac{d}{dr}\biggl(r^2\biggl(1-\frac{r_s}{r}\biggr)\biggr)\frac{d}{dr}R_l(\omega|r)
+\biggl[\frac{\omega^2r^2}{c^2(1-r_s/r)}-l(l+1)\biggr]R_l(\omega|r)=0
\label{radial equation}\;.
 \eeq
This equation can also be written as
 \beq
{d^2\/d{r^\ast}^2} (rR_l(\omega|r))+[\omega^2-V_l(r)](rR_l(\omega|r))=0\label{changed radial function}
 \eeq
with
 \beq
V_l(r)=\biggl(1-{2M\/r}\biggr)\biggl[{l(l+1)\/r^2}+{2M\/r^3}\biggr]\label{effective potential}
 \eeq
being the effective potential. Eq.~(\ref{changed radial function}) admits the following asymptotic solutions, since $V_l(r)\sim0$ in the asymptotic regions ($r\sim2M$ and $r\rightarrow\infty$)
 \bea
 \label{asymp1}
&&\vec{R}_l(\omega|r)\sim\left\{
                    \begin{aligned}
                 &r^{-1}e^{i\omega
r_\ast/c}+\vec{A}_l(\omega)r^{-1}e^{-i\omega r_\ast/c},\;\; r
\rightarrow r_s\;,\cr
                  &
                  {B}_l(\omega)r^{-1}e^{i\omega r_\ast/c},\;\;\quad\quad \quad\quad\quad\;\;\;r
\rightarrow\infty\;,\cr
                          \end{aligned} \right.\\
  \label{asymp2}
&&\cev{R_l}(\omega|r)\sim\left\{
                      \begin{aligned}
&{B}_l(\omega)r^{-1}e^{-i\omega r_\ast/c},\;\;\quad\quad \quad\quad
\quad\;r \rightarrow r_s\;,\cr &r^{-1}e^{-i\omega
r_\ast/c}+\cev{A_l}(\omega)r^{-1}e^{i\omega r_\ast/c},\;\;r
\rightarrow \infty\;.
                       \end{aligned} \right.
 \eea
Here $r_\ast=r+r_s\ln(\frac{r}{r_s}-1)$ is the Regge-Wheeler tortoise coordinate, and ${B}_l(\omega)$
and ${A_l}(\omega)$ are respectively the transmission and reflection coefficients, which obey
 \bea
&|\vec{A}_l(\omega)|=|\cev{A}_l(\omega)|\;,&\\
&1-|\vec{A}_l(\omega)|^2=1-|\cev{A}_l(\omega)|^2=|B_l(\omega)|^2\;.&
 \eea
The physical interpretation of these modes is that $\vec u$
represents the modes emerging from the past horizon and $\cev u$ denotes
those coming in from infinity. The Boulware vacuum state is defined
by requiring normal modes to be positive frequency with respect to
the Killing vector $\partial/\partial t$ with respect to which the
exterior region is static \cite{Boulware75}.

The field operator can be expanded in terms of the complete set of
the basis modes, and the Feynamn propagator (for $t>t'$) of the
massless scalar field in the vacuum is given
by~\cite{Fulling77,Candelas80}
 \bea
D_B^+(x,x')=\frac{\hbar}{4\pi c}\sum_{lm}|Y_{lm}(\theta,\varphi)|^2\,
    \int_{0}^{+\infty}\frac{d\omega}{\omega}
    e^{-i\omega\Delta t}[\,|\vec{R}_l(\omega|r)|^2
    +|\cev{R}_l(\omega|r)|^2].\label{Boulware propogator}
 \eea
It then readily follows that
$C^{F}(x(\tau),x(\tau'))=\frac{1}{2}(D_B^+(x,x')+D_B^+(x',x))$ and
$\chi^{F}(x(\tau),x(\tau'))=\frac{1}{2}(D_B^+(x,x')-D_B^+(x',x))$.
Inserting these statistical functions into Eqs.~(\ref{Evf}) and
(\ref{Err}), we can separately calculate  the contributions of
vacuum fluctuations and radiation reaction to the shift of an energy
level of the atom. Adding them up, we obtain the total shift of
level $b$
 \bea
\delta E_b=-\frac{\mu^2}{16\pi^2c}\sum_{d}|\langle
b|S_2(0)|d\rangle|^2\int_{0}^{\infty}d\omega\;
          \frac{\omega}{\frac{\omega}{\sqrt{g_{00}}}-\omega_{bd}}\times[\vec{\textsf{g}}(\omega|r)+\cev{\textsf{g}}(\omega|r)]\;,
 \eea
where
 \bea
\vec{\textsf{g}}(\omega|r)=\frac{1}{\omega^2}\sum_{l=0}^{\infty}(2l+1)|\vec{R}_l(\omega|r)|^2\;,\nonumber\\
\cev{\textsf{g}}(\omega|r)=\frac{1}{\omega^2}\sum_{l=0}^{\infty}(2l+1)|\cev{R}_l(\omega|r)|^2\;,
 \eea
$g_{00}=1-r_s/r$ and $\sum_{d}|\langle b|S_2(0)|d\rangle|^2=1/4$.

The Lamb shift of the atom is obtained by subtracting the shift of the ground state
from that of the excited state
 \bea
\Delta_B&=&-\frac{\mu^2}{64\pi^2c}
          \int_{0}^{\infty}\frac{d\omega}{\omega}\;
          \biggl(\frac{1}{\frac{\omega}{\sqrt{g_{00}}}-\omega_0}
          -\frac{1}{\frac{\omega}{\sqrt{g_{00}}}+\omega_0}\biggr)\times
          \textsf{g}_s(\omega|r)
          \label{Boulware LS}
 \eea
with
 \beq
\textsf{g}_s(\omega|r)=\sum_{l=0}^{\infty}(2l+1)|\vec{R}_l(\omega|r)|^2+\sum_{l=0}^{\infty}(2l+1)|\cev{R}_l(\omega|r)|^2\;.\label{gs}
 \eeq
Since we are interested in the correction to the shift due to 
nonzero spacetime curvature, we will compare $\Delta_B$ to the Lamb
shift in  flat spacetime $\Delta_M$, which is given, in the same
notation, by~\cite{Audretsch95}
 \bea
\Delta_M=-\frac{\mu^2}{16\pi^2c^3}
          \int_{0}^{\infty}d\omega\;
          \biggl(\frac{\omega}{\omega-\omega_{0}}
          -\frac{\omega}{\omega+\omega_{0}}\biggr)\;.
          \label{Minkowski LS}
 \eea
To characterize the correction induced, we introduce a ratio between the two
 \beq
F(r)=\frac{\Delta_B}{\Delta_M}\;.\label{re LS}
 \eeq
Here, both $\Delta_B$ and $\Delta_M$ are formally divergent. However, the divergence can be dealt with by introducing a cut-off factor~\cite{Bethe47,Welton} or resorting to a fully relativistic approach where no cut-off is present~\cite{Lamb49, French49}.
We choose the former in the present paper for simplicity.
 In fact, if the cut-off is chosen as the electron mass as suggested by Bethe~\cite{Bethe47}, then the results from  the two methods agree. As a result,
  we expect the ratio $F(r)$ to be cut-off independent.
 To calculate the relative correction, $F(r)$, at any given $r$, we need to know function $\textsf{g}_s(\omega|r)$ which depends on the radial parts, $\vec{R}_l(\omega|r)$ and $\cev{R}_l(\omega|r)$, of the scalar field. Analytic solutions to the radial equation
Eq.~(\ref{radial equation}) are however hard to find, so we now resort to a numerical approach~\cite{Jenson89, Jenson92}. In this approach, one expands the radial functions ${R}_l(\omega|r)$ as an infinite power series of $(1-r_s/r)$ at a given position $r$ outside the massive body
 \bea
\left\{
  \begin{array}{ll}
    \vec{R}_l(\omega|\;r)=\frac{1}{r_s}\;e^{i\omega
r_\ast/c}S^*_{\omega l}(r)+\frac{\vec{A}_l(\omega)}{r_s}\;
    e^{-i\omega r_\ast/c}S_{\omega l}(r),  \\
    \cev{R_l}(\omega|\;r)=\frac{B_l(\omega)}{r_s}\;e^{-i\omega r_\ast/c}S_{\omega l}(r),
  \end{array}
\right.
  \label{numerical radial function}
 \eea
where
 \beq
S_{\omega l}(r)=\sum_{k=0}^{\infty}a_k(l,\omega)\biggl(1-\frac{r_s}{r}\biggr)^k\;.\label{general S}
 \eeq
Inserting Eq.~(\ref{numerical radial function}) into
Eq.~(\ref{radial equation}), we obtain, after some simplifications,
the recursive relation of the coefficients
 \bea
&&k\biggl(k-2ir_s\frac{\omega}{c}\biggr)a_k(l,\omega)+\biggl[-3(k-1)^2+2i(k-1)r_s\frac{\omega}{c}-2ir_s\frac{\omega}{c}-l(l+1)\biggr]\times
\nonumber\\&&\quad
 a_{k-1}(l,\omega)+[3(k-2)^2+l(l+1)]a_{k-2}(l,\omega)-(k-3)^2a_{k-3}(l,\omega)=0
 \eea
with $a_0(l,\omega)=1$ and $a_k(l,\omega)=0$ for $k<0$. For any given $\omega$ and $l$,  coefficients $a_k(l,\omega)$
for any $k>0$ can be found using the above relation.

\begin{figure}[htbp]
\centering
\includegraphics[scale=0.8]{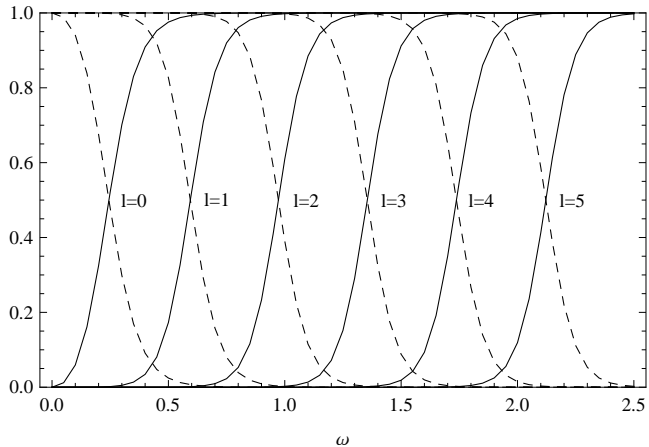}
\caption{$|\protect\vec{A}_l(\omega)|^2$ and $|B_l(\omega)|^2$ as
functions of $\omega$ for $l=0,1,2,3,4,5$. The horizontal axis
is in the unit of $c/r_s$. The dashed lines represent
$|\protect\vec{A}_l(\omega)|^2$ and the solid $|B_l(\omega)|^2$. }
\label{R-T}
\end{figure}

Now our task is to evaluate the transmission and reflection coefficients, ${B}_l(\omega)$ and $\vec{A}_l(\omega)$,
which can be determined by comparing the series representation of the radial functions (\ref{numerical radial function})
with their asymptotic forms (\ref{asymp1}) and (\ref{asymp2}) at large radii. For the radial function of the outgoing modes
$\vec{R}_l(\omega|r)$, one has, at infinity, i.e., $r\rightarrow\infty$,
 \beq
\frac{1}{r_s}\;e^{i\omega r_\ast/c}S^*_{\omega l}(r)+\frac{\vec{A}_l(\omega)}{r_s}\;e^{-i\omega r_\ast/c}S_{\omega l}(r)
={B}_l(\omega)\;e^{i\omega r_\ast/c}\biggl(\frac{1}{r}+O\biggl(\frac{1}{r^2}\biggr)\biggr)\;.
 \eeq
After applying the operator
 \beq
\textbf{D}\equiv\frac{d}{dr_*}-i\frac{\omega}{c}+\frac{r-r_s}{r^2}
 \eeq
to both sides of the above equation, the right-hand side becomes $O(r^{-3})$. Then the reflection coefficient is found
  to approximately be
 \beq
\vec{A}_l(\omega)\approx-\frac{e^{2i\omega r_*/c}(\textbf{D}
+i\frac{\omega}{c})S_{\omega l}^*(r)}{(\textbf{D}-i\frac{\omega}{c})S_{\omega l}(r)}\;.\label{reflection coefficient}
 \eeq
Similar operations on the radial function of the ingoing modes  $\cev{R_l}(\omega|\;r)$ yield the
transmission coefficients
 \beq
B_l(\omega)\approx-\frac{2ir_s\frac{\omega}{c}}{r(\textbf{D}-i\frac{\omega}{c})S_{\omega l}(r)}\;.\label{transmission coefficient}
 \eeq
After simplifications, both coefficients can be expressed in terms of an infinite  series with
respect to $a_k(l,\omega)$ and $(1-r_s/r)$ as
 \bea
\vec{A}_l(\omega)&=&-e^{2ir_s\frac{\omega}{c}[\frac{r}{r_s}+ln(\frac{r}{r_s}-1)]}\nonumber\\&&\times
\frac{\sum_{k=0}^{\infty}[\frac{r_s^2}{2r^2}(k-1)+\frac{r_s}{2r}]a_k^*(l,\omega)(1-\frac{r_s}{r})^k}
{\sum_{k=0}^{\infty}[\frac{r_s^2}{2r^2}(k-1)+\frac{r_s}{2r}-ir_s\frac{\omega}{c}]a_k(l,\omega)(1-\frac{r_s}{r})^k}
 \eea
and
 \bea
B_l(\omega)=-\frac{ir_s\frac{\omega}{c}}{\sum_{k=0}^{\infty}
[\frac{r_s}{2r}k+\frac{1}{2}(1-\frac{r_s}{r})-ir\frac{\omega}{c}]a_k(l,\omega)(1-\frac{r_s}{r})^k}\;.
 \eea
In our numerical computation, these coefficients are evaluated for large and increasing $r$ until they have converged to a desired accuracy such that $|1-(|\vec{A}_l(\omega)|^2+|B_l(\omega)|^2)|$ is less than a prescribed number $A_1$. Fig.~\ref{R-T} shows the numerically obtained values of $|\vec{A}_l(\omega)|^2$ and $|B_l(\omega)|^2$ for various $l$ and $\omega$ for
$r=5\times10^3\;r_s$ and
$A_1=10^{-6}$. It is obvious that both coefficients change very rapidly with $\omega$.  On one hand, our numerical result on the transmission coefficient is consistent with what one expects from a geometrical approximation, i.e., for a given quantum number $l$, $|B_l(\omega)|^2\sim1$ at high frequencies \cite{Dewitt75}, and on the other hand, it also agrees well with the approximate analytical expressions in the low frequency limit given by Page in Ref.~\cite{Page}.

\begin{figure}[htbp]
\centering
\includegraphics[scale=0.8]{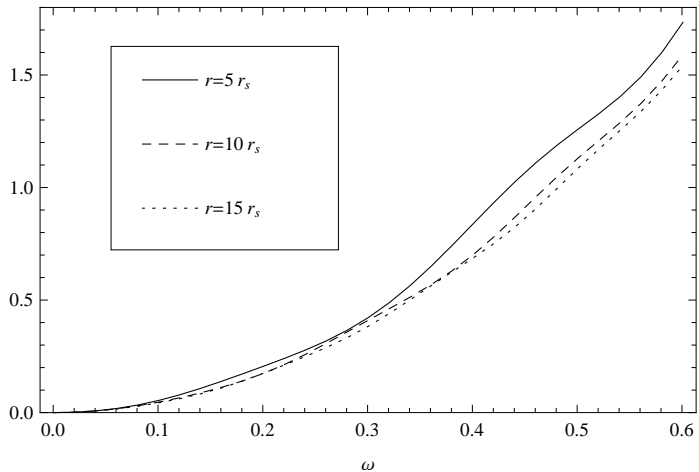}
\caption{ $r_s^{2}\; \protect \textsf{g}_s(\omega|r)$ as function of
$\omega$, $r=5\;r_s,10\;r_s,15\;r_s$. The horizontal axis is
in the unit of $c/r_s$. } \label{g_s}
\end{figure}

Similarly, we can numerically evaluate $S_{\omega l}(r)$ for various
$\omega$ and $l$ that range from zero to some large values such that
a desired accuracy $A_2$ is achieved. In this paper, we terminate
the summation in $S_{\omega l}(r)$ for a given $r$, when
$\biggl|\frac{a_{k_1}(l,\omega)(1-r_s/r)^{k_1}}{\sum_{k=0}^{k_1}a_k(l,\omega)(1-r_s/r)^k}\biggr|\leq
10^{-10}$. Now with $S_{\omega l}(r)$
and the reflection and transmission coefficients given, we can
compute $\textsf{g}_s(\omega|r)$, which is needed in our evaluation
of $\Delta_B$, by appealing to Eqs.~(\ref{numerical radial
function}) and (\ref{general S}). In the numerical computation of  $\textsf{g}_s(\omega|r)$, we
have set the accuracy to terminate the summation with respect to the quantum number $l$ to be $A_1=10^{-6}$
such that $\frac{(2l_1+1)|R_{l_1}(\omega|r)|^2}{\sum_{l=0}^{l_1}(2l+1)|R_{l}(\omega|r)|^2}\leq
A_1$ for both the outgoing and ingoing modes. Fig.~\ref{g_s} shows how $\textsf{g}_s(\omega|r)$ changes with $\omega$. Notice that $r_s^{2}\;\textsf{g}_s(\omega|r)$ is plotted instead of
$\textsf{g}_s(\omega|r)$ for convenience.  One can see from Fig.~\ref{g_s} that
$\textsf{g}_s(\omega|r)$ is a monotonously increasing function of $\omega$.

\begin{figure}[!htb]
\centering
\includegraphics[scale=0.8]{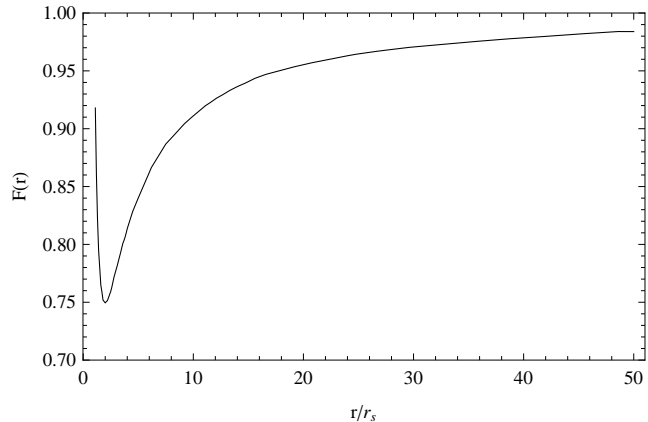}
\caption{The relative correction to the Lamb shift of a static atom
at different radial position, $r$, which ranges from $1.1\;r_s$ to
$50\;r_s$. } \label{relative-LS}
\end{figure}

The relative correction to the Lamb shift at different radii can now
be obtained by numerically integrating both the denominator and the
numerator in Eq.~(\ref{re LS}).
Let us note that both the Lamb shift
of an inertial atom in Minkowski spacetime $\Delta_M$ and that of a
static one outside a massive object $\Delta_B$ are formally
divergent. As we have already mentioned, we deal with the divergence
by introducing a cut-off.
For $\Delta_M$, the cut-off factor is chosen as ${m_ec^2\/\hbar}$ where $m_e$ is the mass of the electron~\cite{Welton,Bethe47}. However, for $\Delta_B$, a red-shift
factor should be included. 
The reason is as follows.  Our calculation of the Lamb shift involves the  proper time of the atom (refer to  Eqs.~(\ref{Evf}) and (\ref{Err})) but the field modes, Eqs.~(\ref{outgoing modes}) and (\ref{ingoing modes}), are expressed in terms of the Schwarzschild coordinate time. Therefore, the cut-off in $\Delta_B$ should be chosen as $\sqrt{g_{00}}\;{m_ec^2\/\hbar}$, since the frequency measured in the Schwarzschild coordinate  $\omega_s$, and that in the  proper frame of the atom  $\omega_p$ are related by  $\omega_s=\omega_p\sqrt{g_{00}}$.
Taking the energy-level spacing of the atom to be $\omega_0=1.271\times10^{16}s^{-1}$ which is comparable to the transition frequency of a hydrogen atom and is within the frequency range of visible light, we numerically calculate the relative correction to the Lamb shift of a two-level atom fixed at various radial distances in the unit of the Schwarzschild
radius $r_s$.
We have computed cases corresponding to various masses of the massive body that range from $~1.59\times10^4\sim10^{14}kg$ and our result, which turns out to be independent of the mass of the massive body, is plotted in Fig.~\ref{relative-LS}. This dependence should not come as a surprise since the  radial distance is measured in terms of the Schwarzschild radius $r_s$ which varies with the mass of the object.

A few comments are now in order. First, the relative correction $F(r)$ is always smaller than unity. This means that at any position, the Lamb shift of the atom at a fixed radial distance outside a massive object is always smaller than that of an inertial one in flat spacetime.
It is worth pointing out that this result seems incompatible with the analytical approximation $\Delta_B\approx[1+f(r)]\Delta_M$ previously obtained~\cite{ZhouYu} (see Eq.~(39) in Ref.~\cite{ZhouYu}) where $f(r)$ is positive. This discrepancy is caused by the
failure of estimating, in the asymptotic regions, the summation over the ingoing and outgoing modes, $\vec{R}_l(\omega|r)$, $\cev{R}_l(\omega|r)$, to the same order of approximation in
Eq.~(29) and Eq.~(30) in Ref.~\cite{ZhouYu} which are actually taken from  Ref.~\cite{Candelas80}. In fact, to keep the approximation to the same order, terms which are of the order next to
$\frac{4\omega^2}{(1-2M/r)}$ should be added both to Eq.~(29) and Eq.~(30) in Ref.~\cite{ZhouYu} (also to corresponding equations in Ref.~\cite{Candelas80}). However, the analytical expressions for these terms are unfortunately hard to find. The numerical calculations we have performed here show that their contributions are actually negative and overtake $f(r)$ so that the net effect is to make  the Lamb shift smaller. As a result, $f(r)$ in Eq.~(33) in Ref.~\cite{ZhouYu} should be replaced by a negative-valued function. We also want to point out that the Lamb shift we are calculating here is in close analogy with that near a conducting surface in a flat spacetime~\cite{CP,ZY}. In a flat spacetime with a conducting surface, the vacuum field modes are completely reflected at the boundary, whereas, in the present case, they are partially reflected and this reflection occurs at every point in space in contrast to the flat spacetime case where the reflection happens only at the boundary. Second, near the horizon, the Lamb shift of the atom decreases rapidly as the radial distance increases until it reaches the minimal value at $\sim 2.0\;r_s$ where the correction is about $25\%$, then it grows very fast, but the slope flattens up at about $20\;r_s$ where the correction is still as large as $4.8\%$. After that the Lamb shift increases very slowly and is expected to approach the value in the Minkowski spacetime at infinity where the spacetime is asymptotically flat. Remarkably, even for a radial distance as far as $50\;r_s$, the correction is still as large as about $1.6\%$. Finally, let us note that $r=r_s$ (where the horizon is located) is a singular point, so in our numerical calculations, we can not start exactly at $r=r_s$ but rather at a point close to it and $r=1.1r_s$ in Fig.~\ref{relative-LS} is just such a point. The closer this point is to the horizon, the closer the Lamb shift is to the value in the Minkowski
spacetime.
This is consistent with our previous result that the Lamb shift of the atom at the horizon approaches the value in the Minkowski spacetime~\cite{ZhouYu}. As a matter of fact, the disappearance of the correction to the Lamb shift close to the horizon is a reflection of the fact that  the effective potential, Eq.~(\ref{effective potential}), which characterizes the scattering of the vacuum field modes off the spacetime curvature, diminishes to zero near the horizon (and at spatial infinity). This point is also discussed in Ref.~\cite{ZhouYu}.

So, for a very compact super-massive astrophysical body such as a
neutron star or a  super-massive black hole in the active galactic
nuclei or even a stellar mass black hole, the curvature-induced
correction to the Lamb shift is remarkably large. In principle, we
can look at the spectra  from a distant compact super-massive body
to find such corrections. Therefore, our results suggest an
interesting way to test fundamental quantum effects using
astrophysical observations. In terms of the possible observation
tests, one should keep in mind that our result is the Lamb shift as
seen by a static observer at the position of the atom.  For
observation, the physically significant quantities are the ones
related to the observations performed at a very large radial distance from
the center of gravity. Let us note that the fact that the Lamb shift
as observed by a distant observer will be gravitationally red-shifted
 makes the correction seem larger than it
actually is. Table.~\ref{tab:near 2M} and Table.~\ref{tab:large r}
show the relative correction to the Lamb shift near the horizon
$r\sim r_s$ and at larger $r$ respectively. Notice that  $F(r)$ and
$F'(r)$ represent, respectively, the relative Lamb shift observed by
a static observer at the position of the atom and that by a distant
one at the spatial infinity.

\begin{table}[h]
\centering \caption{The relative correction to the Lamb shift near
$r\sim r_s$.}\label{tab:near 2M}
\begin{tabular}{|c|c|c|c|c|c|c|c|c|c|c|}
  \hline
  $r/r_s$ & 1.1 & 1.3 & 1.5 & 1.7 & 1.9 & 2.1 & 2.3 & 2.5 & 2.7 & 2.9 \\
  \hline
  $F(r)$ & 0.918 & 0.821 & 0.780 & 0.759 & 0.751 & 0.752 & 0.756 & 0.759 & 0.769 & 0.777\\
  \hline
  $F'(r)$& 0.277 &0.394 & 0.450 & 0.487 & 0.517 & 0.544 & 0.568 & 0.588 & 0.610 & 0.629\\
  \hline
\end{tabular}

\end{table}
\begin{table}[h]
\centering \caption{The relative correction to the Lamb shift at
larger $r$.}\label{tab:large r}
\begin{tabular}{|c|c|c|c|c|c|c|c|c|c|c|}
  \hline
  $r/r_s$ & 5 & 10 & 15 & 20 & 25 & 30 & 35 & 40 & 45 & 50 \\
  \hline
  $F(r)$ & 0.840 & 0.909 & 0.939 & 0.952 & 0.964 & 0.967 & 0.976 & 0.980 & 0.983 & 0.984 \\
  \hline
  $F'(r)$& 0.751 & 0.862 & 0.907 & 0.928 & 0.945 & 0.951 & 0.962 & 0.968 & 0.972 & 0.974\\
  \hline
\end{tabular}
\end{table}

Finally, let us point out that the correction to the Lamb shift we have just calculated is for an atom on a  trajectory at a fixed radial distance outside a compact massive body.  Now a question arises as to how the trajectory is realized, in other words, who is holding the atom, and this question is specially relevant when actual observational verifications are desired.  A natural way to keep an atom  at a fixed radial distance is to let it undergo circular geodesic motion. It has been shown that bound circular orbits are possible for massive particles only when $r\ge4GM/c^2$ and among them only orbits with $r\geq6GM/c^2$ are stable~\cite{Bardeen}. The closer is the orbit to $r=4GM/c^2$, the greater is the speed of the particle.  However, an atom under geodesic circular motion will be subjected to the circular Unruh effect which has already been studied in the literature~\cite{Letaw,Bell,Takagi,Audretsch} and as such  additional corrections to the Lamb shift will 
result. However, as we will demonstrate next,  such corrections are in general much smaller than the curvature induced corrections for a typical very compact massive body such as a neutron star or a black hole. To this end, let us note that for a two-level atom in uniform circular motion, the correction to the Lamb shift is given  in the high-velocity limit, $v/c\gtrsim0.85$, by~\cite{Audretsch}
 \beq
\Delta_{C}={\mu^2a A\/32\sqrt{3}\pi^2c^4}
\biggl(e^{-2\sqrt{3}B\omega_0c/a}\overline{\mathrm{Ei}}(2\sqrt{3}B\omega_0c/a)
-e^{2\sqrt{3}B\omega_0c/a}\overline{\mathrm{Ei}}(-2\sqrt{3}B\omega_0c/a)\biggr)\;,
 \eeq
where $\overline{\mathrm{Ei}}$ denotes the principal value of the exponential integral function \cite{Gradshteyn},
$a$ is the centripetal acceleration, $\omega_0$ the energy gap between the two levels of the atom, $A=1+{3\/5}({c\/v\gamma})^{2}$, $B=1-{1\/5}({c\/v\gamma})^{2}$ with $\gamma=(1-{v^2\/c^2})^{-1/2}$ and $v$ is the
 velocity of the atom. Notice that here we have corrected a couple of typos in the original formula given in Ref.~\cite{Audretsch}.  To estimate how large ${\Delta_C}$ typically will be, let us  take a neutron star of a solar mass $M\sim10^{30}kg$ as an example. If we choose  $r=4.2GM/c^2$ which is very close to the inner most circular orbit and $\omega_0\sim10^{16}s^{-1}$ which is the typical hydrogen transition frequency, then we can easily calculate the  velocity  using the formula in Ref.~\cite{Bardeen} to get $v\approx0.913c$  and this leads to the centripetal acceleration $a={v^2\gamma^2\/r}\approx1.448\times10^{14}m/s^2$, $A\approx1.120$ and $B\approx0.960$. With these constants given, we can evaluate $\Delta_{C}$ and compare it with $\Delta_{M}$ to find ${\Delta_C\/\Delta_M}\sim10^{-23}$. This ratio becomes even smaller when the radius of the circular orbit is larger. This shows that the corrections due to the circular motion to keep the atom at a fixed radial distance outside very compact massive objects is completely negligible as compared to those caused by the spacetime curvature. So, in practical sense, our results also apply to atoms orbiting a very compact massive astronomical object in circular motion at a given radial distance.

We are grateful to Tingyun Shi, Lingyan Tang, Zhengxiang Zhong and
Jiawei Hu for useful suggestions on numerical calculations, and to
Ziqing Xie and You Zou for providing computing resources at the Key
Laboratory of Computational and Stochastic Mathematics and Its
Applications. This work was supported in part by the NNSFC under
Grants No. 11075083, and No. 10935013; the Zhejiang Provincial
Natural Science Foundation of China under Grant No. Z6100077; the
National Basic Research Program of China under Grant No.
2010CB832803; the PCSIRT under Grant No. IRT0964; the Hunan
Provincial Natural Science Foundation of China under Grant No.
11JJ7001; and Hunan Provincial Innovation Foundation For
Postgraduate under Grant No. CX2011B187.


\end{document}